\documentclass[runningheads,a4paper]{llncs}

\usepackage{amssymb}
\usepackage{graphicx}

\usepackage{hyperref}
\usepackage{url}
\urldef{\mailsa}\path|dignatov@hse.ru|
\urldef{\mailsb}\path|khvorykh@img.ras.ru|
\newcommand{\keywords}[1]{\par\addvspace\baselineskip
\noindent\keywordname\enspace\ignorespaces#1}
\usepackage[ruled,vlined]{algorithm2e}

\SetKwFor{ParFor}{parallel for}{do}{endfor} 

\usepackage{listliketab}
\usepackage{multirow}

\usepackage{algorithmic}
\usepackage{url}
\usepackage{amsmath}

\begin{document}

\mainmatter  

\title{Object-Attribute Biclustering for Elimination of Missing Genotypes in Ischemic Stroke Genome-Wide Data}

\titlerunning{Object-Attribute Biclustering for Elimination of Missing Genotypes}

%
%
\author{Dmitry I. Ignatov\orcidID{0000-0002-6584-8534}\inst{1}  \and Gennady V. Khvorykh\orcidID{0000-0001-8927-5921}\inst{2} \and Andrey V. Khrunin\orcidID{0000-0002-7848-4688}\inst{2} \and Stefan Nikoli\'c\orcidID{0000-0001-8105-5465}\inst{1} \and Makhmud Shaban\orcidID{0000-0003-4773-8398}\inst{1} \and Elizaveta A. Petrova \inst{3} \and Evgeniya A. Koltsova \inst{3} \and Fouzi Takelait\orcidID{0000-0003-3497-608X}\inst{1} \and Dmitrii Egurnov\orcidID{0000-0002-8195-1670}\inst{1}}

\authorrunning{Dmitry I. Ignatov et al.}


\institute{National Research University Higher School of Economics, Russia\\
\mailsa, \url{http://www.hse.ru}\\
\and
Institute of Molecular Genetics of National Research Centre ``Kurchatov Institute'', Russia\\
\mailsb, \url{http://img.ras.ru}\\
\and
Pirogov Russian National Research Medical University, Russia\\
\url{http://rsmu.ru}
}

%
%

\toctitle{Object-Attribute Biclustering for Elimination of Missing Genotypes in Ischemic Stroke Data}
\maketitle

\begin{abstract}
Missing genotypes can affect the efficacy of machine learning approaches to identify the risk genetic variants of common diseases and traits. The problem occurs when genotypic data are collected from different experiments with different DNA microarrays, each being characterised by its pattern of uncalled (missing) genotypes. This can prevent the machine learning classifier from assigning the classes correctly. To tackle this issue, we used well-developed notions of object-attribute biclusters and formal concepts that correspond to dense subrelations in the binary relation $\textit{patients} \times \textit{SNPs}$. The paper contains experimental results on applying a biclustering algorithm to a large real-world dataset collected for studying the genetic bases of ischemic stroke. The algorithm could identify large dense biclusters in the genotypic matrix for further processing, which in return significantly improved the quality of machine learning classifiers. The proposed algorithm was also able to generate biclusters for the whole dataset without size constraints in comparison to the In-Close4 algorithm for generation of formal concepts.

\end{abstract}

\section{Introduction}
The recent progress in studying different aspects of human health and diversity (e.g., genetics of common diseases and traits, human population structure, and relationships) is associated with the development of high-throughput genotyping technologies, particularly with massive parallel genotyping of Single Nucleotide Polymorphisms (SNPs) by DNA-microarrays~\cite{pmid23288464}. They allowed the determination of hundreds of thousands and millions of SNPs in one experiment and were the basis for conducting genome-wide association studies (GWAS). Although thousands of genetic loci have been revealed in GWAS, there are practical problems with replicating the associations identified in different studies. They seem to be due to both limitations in the methodology of the GWAS approach itself and differences between various studies in data design and analysis~\cite{pmid29876890}. The machine learning (ML) approaches were found to be quite promising in this field~\cite{pmid32351543}. 

Genotyping by microarrays is efficient and cost-effective, but missing data appear. GWAS is based on a comparison of frequencies of genetic variants among patients and healthy people. It assumes that all genotypes are provided (usually, their percentage is defined by a genotype calling threshold). In this article, we demonstrate that missing data can affect not only statistical analysis but also the ML algorithms. The classifiers can fail because of missing values (uncalled genotypes) being distributed non-randomly. We assume that each set of DNA-microarray can possess a specific pattern of missing values marking both the dataset of patients and healthy people. Therefore, the missing data needs to be carefully estimated and processed without dropping too many SNPs that may contain crucial genetic information. 

To overcome the problem of missing data, we aimed to apply a technique capable of discovering some groupings in a dataset by looking at the similarity across all individuals and their genotypes. The raw datasets can be converted into an integer matrix, where individuals are in rows, SNPs are in columns, and cells contain genotypes. For each SNP, the person can have either AA, AB, or BB genotype, where A and B are the alleles. Thus the genotypes can be coded as 0, 1, and 2, representing the counts of allele B.  

The proposed method can simultaneously cluster rows and columns in a data matrix to find homogeneous submatrices~\cite{tanay2002discovering}, which can overlap. Each of these submatrices is called a bicluster~\cite{Mirkin:1996}, and the process of finding them is called biclustering~\cite{tanay2002discovering,madeira2004biclustering,cheng2000biclustering,tanay2005biclustering,busygin2008biclustering}.

Biclustering in genotype data allows identifying sets of individuals sharing SNPs with missing genotypes. A bicluster arises when there is a strong relationship between a specific set of objects and a specific set of attributes in a data table. A particular kind of bicluster if a formal concept in Formal Concept Analysis (FCA)~\cite{Ganter:1999:FCA}. A formal concept is a pair of the form (extent, intent), where extent consists of all objects that share the attributes in intent, and dually the intent consists of all attributes shared by the objects in extent. Formal concepts have a desirable property of being homogeneous and closed in the algebraic sense, which resulted in their extensive use in Gene Expression Analysis (GEA) \cite{Besson:2005,Blachon:2007,Kaytoue:2011,Andrews:2013}.

A concept-based bicluster (or object-attribute bicluster)~\cite{Ignatov:2010} is a scalable approximation of a formal concept withe the following advantages:

\begin{enumerate}
  \item Reduced number of patterns to analyze;
  \item Reduced computational cost (polynomial vs. exponential);
  \item Manual (interactive) tuning of bicluster density threshold;
  \item Tolerance to missing (object, attribute) pairs.
\end{enumerate}

In this paper, we propose an extended biclustering algorithm of~\cite{ignatov2012concept} that can identify large biclusters with missing genotypes for categorical data (many-valued contexts with a selected value). This algorithm can generate a smaller amount of dense object-attribute biclusters than that of existing exact algorithms for formal concepts like concept miner In-Close4~\cite{Andrews:2011}, and is, therefore, better suited for large datasets. Moreover, during experimentation with the ischemic stroke dataset, we found that the number of large dense biclusters identified by our algorithm is significantly lower than the number of formal concepts extracted by In-Close4\footnote{https://sourceforge.net/projects/inclose/} and Concept Explorer (ConExp\footnote{http://conexp.sourceforge.net})~\cite{yevtushenko2000system}. 

The paper is organized as follows. In Section~\ref{def}, we recall basic notions from Formal Concept Analysis and Biclustering. In Section~\ref{model}, we introduce a method of FCA-based biclustering and its variants along with bicluster post-processing schemes, consider discussing the complexity of the proposed algorithm.   In Section~\ref{data}, we describe a dataset that consists of a sample of patients and their SNPs collected from various (independent) groups of patients. Then we present the results obtained during experiments on this dataset in Section~\ref{exp} and mention the used hardware and software configuration. Section~\ref{con} concludes the paper.

\section{Basic notions}\label{def}

\subsection{Formal Concept Analysis}\label{FCA}
\begin{definition}
A \textbf{formal context} in FCA~\cite{Ganter:1999:FCA} is a triple $\mathbb{K} = (G, M, I)$ consisting of two sets, $G$ and $M$, and a binary relation $I \subseteq G \times M$ between $G$ and $M$. The triple can be represented by a cross-table consisting of rows $G$, called {\bf objects}, and columns $M$, called {\bf attributes}, and crosses representing incidence relation $I$. Here, $gIm$ or $(g, m) \in I$ means that the object $g$ {\bf has} the attribute $m$.
\end{definition}

\begin{definition}
For $A \subseteq G$ and  $B \subseteq M$, let
\begin{eqnarray*}
A' \stackrel{\rm def}{=} \{m\in M\mid gIm \ {\rm for\ all}\ g\in A\},\mbox{ and } B' \stackrel{\rm def}{=} \{g\in G\mid gIm \ {\rm for\ all}\ m\in B\}.
\end{eqnarray*}

\noindent These two operators are the {\bf derivation operators} for $\mathbb{K} = (G, M, I)$.
\end{definition}

\begin{proposition}
Let $(G, M, I)$ be a formal context, for subsets $A, A_1, A_2 \subseteq G$ and $B \subseteq M$ we have

\begin{multicols}{2}
    \begin{enumerate}
    \item $A_1 \subseteq A_2$ if $A'_2 \subseteq A'_1$,
    \item $A \subseteq A''$,
    \item $A = A''$ (hence, $ A'''' = A''$),
    \item $(A_1 \cup A_2)' = A'_1 \cap A'_2$,
    \item $A \subseteq B' \Leftrightarrow B \subseteq A' \Leftrightarrow A \times B \subseteq I$.
\end{enumerate}
\end{multicols}

Similar properties hold for subsets of attributes.
\end{proposition}

\begin{definition}
A {\bf closure operator} on set $S$ is a mapping $\varphi: 2^S \rightarrow 2^S$ with the following properties:

\noindent Let $X \subseteq S$, then
\begin{enumerate}
    \item $\varphi(\varphi(X)) = \varphi(X)$ ({\it\textbf{idempotency}}),
    \item $X \subseteq \varphi(X)$ ({\it\textbf{extensity}}),
    \item $X \subseteq Y \Rightarrow \varphi(X) \subseteq \varphi(Y)$ ({\it\textbf{monotonicity}}).
\end{enumerate}

\noindent For a closure operator $\varphi$ the set $\varphi(\varphi(X))$ is called {\bf closure} of $X$, while a subset $X \subseteq S$ is called {\bf closed} if $\varphi(\varphi(X)) = X$.\\
\end{definition}

It is evident from properties of derivation operators that for a formal context  $(G, M, I)$, the operators
$$(\cdot)'': 2^G \rightarrow 2^G \mbox{ and } (\cdot)'': 2^M \rightarrow 2^M$$

\noindent are closure operators.

\begin{definition}
$(A, B)$ is a {\bf formal concept} of formal context $\mathbb{K} = (G, M, I)$ iff
$$A \subseteq B,\ B \subseteq M,\ A' = B,\ {\rm and}\ A = B'\mbox{ .}$$
The sets $A$ and $B$ are called the {\bf extent} and the {\bf intent} of the formal concept $(A, B)$, respectively.
\end{definition}

This definition says that every formal concept has two parts, namely, its extent and intent. It follows an old tradition in philosophical
concept logic, as expressed in the {\it Logic of Port Royal, 1662}~\cite{arnauld1662logique}.

\begin{definition}
The set of all formal concepts $\underline{\mathfrak{B}}(B, M, I)$ is partially ordered, given by relation $\leq_\mathbb{K}$:
$$(A_1, B_1) \leq_\mathbb{K} (A_2, B_2) \iff A_1 \subseteq A_2\ ({\rm dually}\ B_2 \subseteq B_1)$$
$\underline{\mathfrak{B}}(B, M, I)$ is called {\bf concept lattice} of the formal context $\mathbb{K}$.
\end{definition}



In case an object has properties like colour or age the corresponding attributes should have values themselves.

\begin{definition}
A {\bf many-valued context} $(G,M,W,J)$ consists of sets $G$, $M$ and $W$ and a ternary relation $J \subseteq G \times M \times W$ for which it holds that 

$$(g,m,w) \in J \mbox{ and } (g,m,v) \in I \mbox{ imply } w=v \mbox{ .}$$

The elements of $M$ are called {\bf (many-valued) attributes} and those of $W$ {\bf attribute values}.

\end{definition}

Since many-valued attributes can be considered as partial maps from $G$ in $W$, it is convenient to write $m(g)=w$.

\subsection{Biclustering}

In \cite{madeira2004biclustering}, {\it bicluster} is defined as a homogeneous submatrix of an input object-attribute matrix of real values in general.  Consider a dataset as a matrix, $A = (X, Y) \in \mathbb{R}^{n \times m}$, with a set of rows/objects/individuals $X=\{x_1, \ldots, x_n\}$ and set of columns/attributes/SNPs $Y = \{y_1, \ldots, y_m\}$. A submatrix constructed from a subset of rows $I \subseteq X$ and that of columns $J \subseteq Y$ is denoted as $(I, J)$ is called a {\bf bicluster} of $A$~\cite{madeira2004biclustering}. The bicluster should satisfy some specific homogeneity properties, which varies from method to another. 

For instance, for the purpose of this research, we use the following FCA-based definition of a bicluster \cite{ignatovphdthesis2010,Ignatov:2010,ignatov2012concept}. 

\begin{definition} For a formal context $\mathbb{K} = (G, M, I)$ any biset $(A,B) \subseteq I$ with $A\neq\emptyset$ and $B \neq\emptyset$ is called a {\bf bicluster}. 
If $(g, m) \in I$, then the bicluster $(A,B)=(m', g')$ is called an object-attribute or {\bf OA-bicluster} with {\bf density} $\rho (A, B) = \frac{|I \cap (A \times B)|}{|A| \cdot |B|}$.
\end{definition}

The density $\rho(m',g')$ of a bicluster $(m',g')$ is the bicluster quality measure that shows how many non-empty pairs the bicluster contains divided by its size.

Several basic properties of OA-biclusters are below.

\begin{proposition}
\hspace{1pt}
\begin{enumerate}
    \item For any bicluster $(A,B)\subseteq 2^G\times 2^M$ it is true that $0 \leq \rho(A,B)\leq 1$,
    \item OA-bicluster $(m', g')$ is a formal concept iff $\rho = 1$,
    \item If $(A, B)$ is a OA-bicluster, there exists (at least one) its {\bf generating pair} $(g,m) \in A \times B$ such that $(m',g')=(A, B)$,
    \item If $(m', g')$ is a OA-bicluster, then $(g'', g') \leq (m', m'')$.
    \item For every $(g,m)\in I$, $(h,n) \in [g]_M \times [m]_G$\footnote{The equivalence classes are $[g]_M=\{h \mid  h \in G, g'=h'\}$ and  $[m]_G=\{n \mid  n \in M, n'=m'\}$.}, it follows $(m',g')=(n',h')$. 
\end{enumerate}
\end{proposition}

\begin{figure}[ht!]
  \centering

    \includegraphics[scale=0.45]{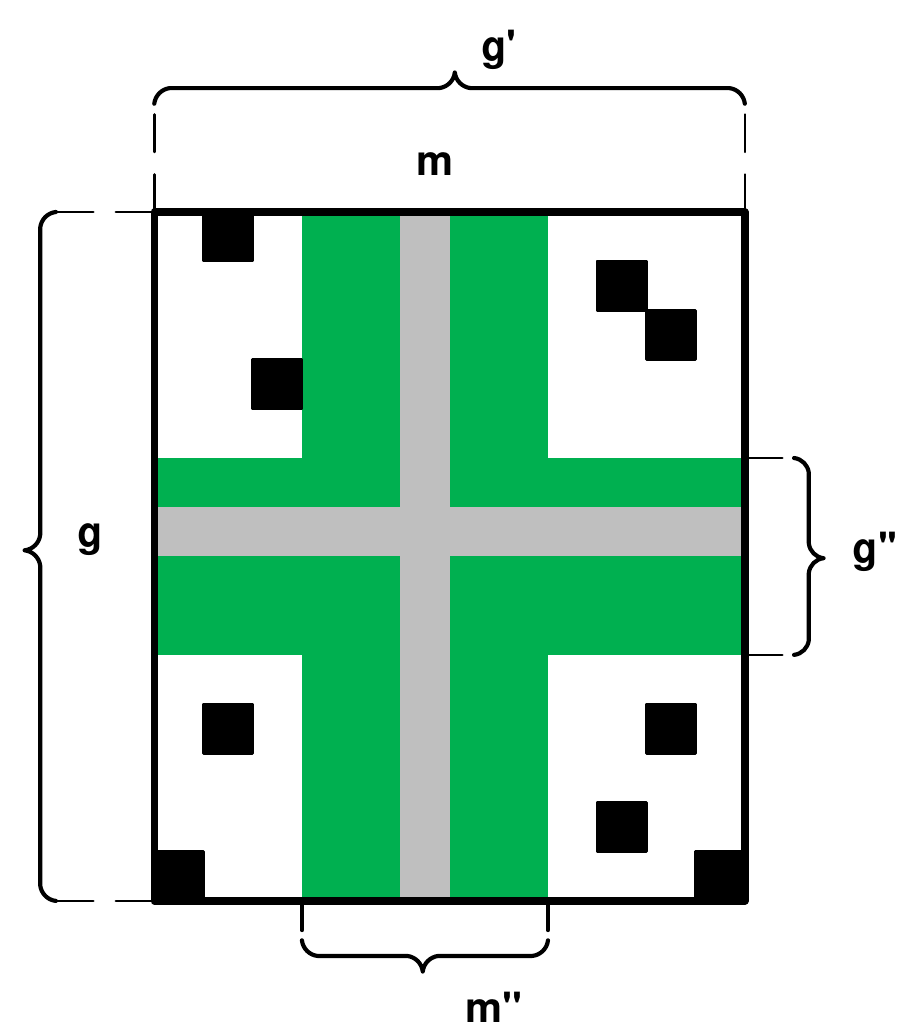}\\

  \caption{{\normalsize OA-bicluster based on object and attribute primes.}}\label{bicl}
\end{figure}

In Fig.~\ref{bicl}, you can see the example of OA-bicluster, for a particular pair $(g, m) \in I$ of a certain context $(G, M, I)$. In general, only the regions $(g'', g')$ and $(m', m'')$ are full of non-empty pairs, i.e. have maximal density $\rho = 1$, since they are object and attribute formal concepts respectively. The black cells indicate non-empty pairs, which one may found in less dense white regions.  

\begin{definition}
Let $(A, B) \in 2^G \times 2^M$ be a OA-bicluster and $\rho_{min} \in (0,1]$, then $(A, B)$ is called dense if it satisfies the constraint $\rho (A, B) \geq \rho_{min}$.
\end{definition}

The number of OA-biclusters of a context can be much less than the number of formal concepts (which may be $2^{\min (|G|,|M|)}$), as stated by the following propositions.

\begin{proposition}
For a formal context $\mathbb{K} = (G, M, I)$ the largest number of OA-biclusters is equal to $|I|$ and all OA-biclusters can be generated in time $\mathcal{O}(|I|)$.
\end{proposition}

\begin{proposition}
For a formal context $\mathbb{K} = (G, M, I)$ and  $\rho_{min} > 0$ the largest number of dense OA-biclusters is equal to $|I|$, all dense OA-biclusters can be generated in time $\mathcal{O}(|I| |G||M|)$.
\end{proposition}

\section{Model and algorithm description}\label{model}
\subsection{Parallel OA-biclustering algorithm}
Algorithm~\ref{Alg:MVBiA} is a straightforward implementation, which takes an initial many-valued formal context and minimal density threshold as parameters and computes dense biclusters for each $(g, m)$as pair in the relation $I$ that indicates which objects have SNP with missing values. However, since OA-biclusters for many-valued contexts were not formally introduced previously, we use a derived formal context with one-valued attributes denoting missing attribute-values of an original genotype matrix to correctly apply the definition of dense OA-bicluster. 

\begin{definition}
Let $\mathbb{K} = (G, M, W, J)$ is a many-valued context and $v \in W$ is a selected value (e.g., denoting the absence of an SNP value), then its {\bf derived context for the value $v$} is $\mathbb{K}_v = (G, M, I)$ where $gIm$ iff $(g,m,v) \in J$.
\end{definition}

For genotype matrices with missing SNP values as many-valued contexts, similar representation can be expressed in terms of co-domains of many-valued attributes (the absence of $m(g)$ means that of the corresponding SNP value)  or by means of nominal scaling with a single attribute for the missing value $v$~\cite{Ganter:1999:FCA}.


If we compare the number of output pattern for formal concepts and dense OA-biclusters, in the worst case these values are $2^{min(|G|, |M|)}$ versus $|I|$. The time complexity of our algorithm is polynomial, $\mathcal{O}(|G||M||I|)$, versus exponential in the worse case for BiMax~\cite{Prelic:2006}, $\mathcal{O}(|G||M||L| \log |L|)$, or $\mathcal{O}(|G|^2|M||L|)$ for \textsc{CbO} algorithms family~\cite{kuznetsov:1996}, where $|L|$ is a number of generated concepts (also considered as biclusters) and is exponential in the worst case $|L| = 2^{min(|G|,|M|)}$.

For calculating biclusters that fulfil a minimum density constraint, we need to perform several steps (see Algorithm~\ref{Alg:MVBiA}). Steps 5-8 consists of applying the Galois operator to all objects in $G$ and steps 9-12 then to all attributes in $M$ within the induced context. The outer for loops are parallel (the concrete implementation may differ), while the internal ones are ordinary for loops. Then all biclusters are enumerated in a parallel manner as well, and only those that fulfil the minimal density requirement are retained (Steps 13-16). Again, efficient implementation of set data-structure for storing biclusters and duplicate elimination of the fly in parallel execution mode are not addressed in the pseudo-code. 

The novelties of this algorithm are that we used parallelization to generate the OA-bicluster giving as input a medium-sized dataset (e.g. $~10^3 \times 10^4$), that is to make our program runs faster, and the possibility to work with selected values reducing many-valued context to contexts with one-valued attributes.

\begin{algorithm}[ht!]
\caption{OA-bicluster generation for a many-valued context.\label{Alg:MVBiA}}
\DontPrintSemicolon
\LinesNumbered
\KwData{$\mathbb K=(G,M,W,J)$ is a many-valued formal context, $\rho_{min}$ is a threshold density value of bicluster density and $v \in W$ is a selected value}
\KwResult{$B=\{(A,B)| (A,B) \mbox{ is an OA-bicluster for value } v \}$}
\Begin{
$Obj.Size:=|G|$\;
$Attr.Size:=|M|$\;
$B \longleftarrow \emptyset$\;
\ParFor{$g \in G$}{
\For{$m \in M$}{
\If{m(g)=v}{
$Obj[g].Add(m)$\;
}
}
}

\ParFor{$m \in M$}{
\For{$g \in G$}{
\If{m(g)=v}{
$Attr[m].Add(g)$\;
}
}
}
\ParFor{$(g,m,w) \in J$}{
\If{w=v}{
\If{$\rho(Attr[m],Obj[g]) \geq \rho_{min}$}{$B := B \cup \{(Attr[m],Obj[g])\}$}

}
}
}
\end{algorithm}

\subsection{One-pass version of the OA-biclustering algorithm}

Let us describe the online problem of finding the set of prime OA-biclusters based on the online OAC-Prime Triclustering~\cite{Gnatyshak:2014}.
Let $\mathbb{K} = (G,M,I)$ be a context.
The user has no a priori knowledge of the elements and even cardinalities of $G$, $M$, and $I$.
At each iteration, we receive a set of pairs (``batch'') from $I$: $J\subseteq I$.
After that, we must process $J$ and get the current version of the set of all biclusters.
It is important in this setting to consider every pair of biclusters different if they have different generating pairs even if their extents and intents are equal, because any other pair can change only one of them, thus making them different.

Also, the algorithm requires that the dictionaries containing the prime sets are implemented as hash-tables or similar efficient key-value structures.
Because of this data structure, the algorithm can efficiently access prime sets.

The algorithm itself is also straightforward (Alg.~\ref{alg:online}).
It takes a set of pairs ($J$) and current versions of the biclusters set ($\mathcal{B}$) and the dictionaries containing prime sets ($PrimesO$ and $PrimesA$) as input and outputs the modified versions of the bicluster set and dictionaries.
The algorithm processes each pair $(g,m)$ of $J$ sequentially (line 1).
On each iteration the algorithm modifies the corresponding prime sets: it adds $m$ to $g^\prime$ (line 2) and $g$ to $m^\prime$ (line 3).

Finally, it adds a new bicluster to the bicluster set.
Note that this bicluster contains pointers to the corresponding prime sets (in the corresponding dictionaries) instead of their copies (line 4).

In effect, this algorithm is very similar to the original OA-biclustering algorithm with some optimizations.
First of all, instead of computing prime sets at the beginning, we modify them on spot, as adding a new pair to the relation modifies only two prime sets by one element.
Secondly, we remove the main loop by using pointers for the bicluster' extents and intents, as we can generate biclusters at the same step as we modify the prime sets.
And third, it uses only one pass through the pairs of the binary relation $I$, instead of enumeration of different pairwise combinations of objects and  attributes.

\begin{algorithm}
\caption{Online generation of OA-biclusters}
\label{alg:online}
    \begin{algorithmic}[1]
    	\REQUIRE $J$~is a set of object-attribute pairs;\\
            $\mathcal{B}=\{\mathbf b=(*X,*Y)\}$~is the current set of OA-biclusters;\\
            $PrimesO$, $PrimesA$;\\
    	\ENSURE $\mathcal{B}=\{\mathbf b=(*X,*Y)\}$;\\
            $PrimesO$, $PrimesA$;
    	\FORALL{$(g,m)\in J$}
    		\STATE $PrimesO[g]:=PrimesO[g]\cup \{m\}$
            \STATE $PrimesA[m]:=PrimesAC[m]\cup \{g\}$
            \STATE $\mathcal{B}:=\mathcal{B}\cup \{(\&PrimesA[m],\&PrimesO[g])\}$
    	\ENDFOR
    \end{algorithmic}
\end{algorithm}

Each step requires constant time: we need to modify two sets and add one bicluster to the set of biclusters.
The total number of steps is equal to $|I|$; the time complexity is linear $\mathcal O(|I|)$. Beside that the algorithm is one-pass.

The memory complexity is the same: for each of $|I|$ steps the size of each dictionary containing prime sets is increased either by one element (if the required prime set is already present), or by one key-value pair (if not).
Since each of these dictionaries requires $\mathcal{O}(|I|)$ memory, the memory complexity is also linear.

\subsection{Post-processing constraints}

Another important step, in addition to this algorithm, is post-processing.
Thus, we may want to remove additional biclusters with the same extent and intent from the output.
Simple constraints like minimal support condition can be processed during this step without increasing the original complexity.
It should be done only during the post-processing step, as the addition of a pair in the main algorithm can change the set of biclusters, and, respectively, the values used to check the conditions.
Finally, if we need to fulfil more difficult constraints like minimal density condition, the time complexity of the post-processing will be higher than that of the original algorithm, but it can be efficiently implemented.

To remove the same biclusters we need to use an efficient hashing procedure that can be improved by implementing it in the main algorithm.
For this, for all prime sets, we need to keep their hash-values with them in the memory.
And finally, when using hash-functions other than LSH function (Locality-Sensitive Hashing)~\cite{Leskovec:2020}, we can calculate hash-values of prime sets as some function of their elements (for example, exclusive disjunction or sum).
Then, when we modify prime sets, we just need to get the result of this function and the new element.
In this case, the hash-value of the bicluster can be calculated as the same function of the hash-values of its extent and intent.

Then it would be enough to implement the bicluster set as a hash-set in order to efficiently remove the additional entries of the same bicluster.


%

Pseudo-code for the basic post-processing (Alg.~\ref{alg:online-post}).

\begin{algorithm}
\caption{Post-processing for the online OA-biclustering algorithm.}
\label{alg:online-post}
    \begin{algorithmic}[1]
    	\REQUIRE $\mathcal{B}=\{\mathbf b=(*X,*Y)\}$~is a full set of biclusters;\\
    	\ENSURE $\overline{\mathcal{B}}=\{\mathbf b=(*X,*Y)\}$~is a processed hash-set of biclusters;
    	\FORALL{$\mathbf b\in \mathcal{B}$}
    		\STATE Calculate $hash(\mathbf b)$
            \IF{$hash(\mathbf b)\not\in \overline{\mathcal{B}}$}
                \STATE $\overline{\mathcal{B}}:=\overline{\mathcal{B}}\cup \{\mathbf b\}$
            \ENDIF
    	\ENDFOR
    \end{algorithmic}
\end{algorithm}

If the names (codes) of the objects and attributes are small enough (the time complexity of computing their hash values is $\mathcal O(1)$), the time complexity of the post-processing is $\mathcal O(|I|)$ if we do not need to calculate densities, and $\mathcal O(|I||G||M|)$ otherwise.
Also, the basic version of the post-processing does not require any additional memory; so, its memory complexity is $O(1)$.

Finally, the algorithm can be easily paralleled by splitting the subset of input pairs into several subsets, processing each of them independently, and merging the resulting sets afterwards, which may lead to distributed computing schemes for larger datasets (cf. \cite{Ignatov:2019}).

In case the output of the post-processing step is stored in a relational database along with the computed statistics and generating pairs, further usage of selection operators~\cite{Codd:1970} is convenient to consider only a specific subset of biclusters.

We use the following operator resulting in a specific subset of biclusters
$$\sigma_{(\alpha_{min} \leq |A| \leq \alpha_{max}) \wedge (\beta_{min} \leq  |B| \leq \beta_{max}) \wedge (\rho_{min} \leq \rho(A,B) \leq \rho_{max})}(\mathcal B), $$

where $|A|$ is the extent size, $|B|$ is the intent size, and $\rho(A,B)$ is the density of OA-bicluster $\mathbf b \in \mathcal B$, respectively. One more reason to use postprocessing is neither monotonic nor anti-monotonic character of the minimal density constraint in the sense of constraints pushing in pattern mining~\cite{Besson:2005,ignatov2012concept}.

\section{Data collection}\label{data}

Collection of patients with ischemic stroke and their clinical characterisation were made at the Pirogov Russian National Research Medical University. The DNA extraction and genotyping of the samples were described previously~\cite{pmid22677768}.

The dataset contains samples  corresponding to individuals with a genetic portrait for each and a group label. The former represents the genotypes determined at many SNPs all over the genome. The latter takes values 0 or 1 depending on whether a person did not have or had a stroke. Each SNP is a vector that components can take values from $\{0, 1, 2, -1\}$, where 0, 1, and 2 denote the genotypes, and -1 indicates a missing value. 

We represent the dataset as a many-valued formal context. In the derived context $\mathbb K=(G,M,I)$, where objects from $G$ stand for samples and attributes from $M$ stand for SNPs, $gIm$ means that an individual $g$ has a missing SNP $m$.  The context has the following parameters $|G| = 1,323$, $|M| = 85,142$, and $|I| = 45,075$ which represents the total number of attributes with missing values in the dataset and cover 0.491\% of the whole data matrix. The number of attributes without missing values is 40,067.

\begin{figure}[ht!]
    \centering
    \includegraphics[width=\textwidth]{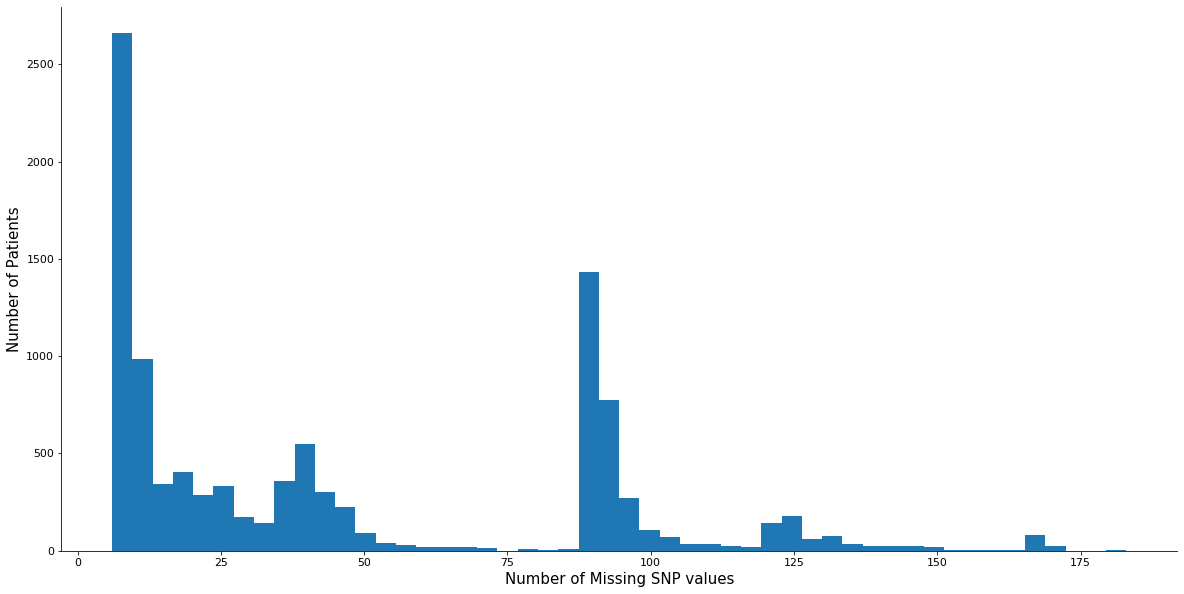} \\
    \caption{{\normalsize The distribution of the number of missing SNP values by columns \textbf{before} elimination.}}\label{fig:distmis}
\end{figure}

The genotypic data were obtained with DNA-microarrays. The dataset was compiled from several experiments where different types of microarrays were applied.
Not all genotypes are equally measured during the experiment. Thus, there is a certain instrumental error. The quality of DNA can also affect the output of the experiments. Fig.~\ref{fig:distmis} shows how many individuals have exactly $N$ missing genotypes per SNP in the dataset.

For instance, many individuals have about 85 missing genotypes per SNP.

\section{Experiments}\label{exp}

\subsection{Hardware and software configuration}\label{hardware}
The experimental results with OA-biclustering generation and processing were obtained on an Intel(R) Core(TM) i5-8265U CPU @ 1.80 GHz with 8 GB of RAM and 64-bit Windows 10 Pro operating system. We used the following software releases to perform our experiments: Python 3.7.4 and Conda 4.8.2.

\subsection{Identification of Biclusters with Missing SNPs}

The following experiment was performed with ischemic stroke data collection: first of all, 383,733 OA-biclusters, with duplicates, were generated after applying the parallel biclustering algorithm to the dataset. 

\begin{figure}[ht!]
  \centering

    \includegraphics[width=1\textwidth]{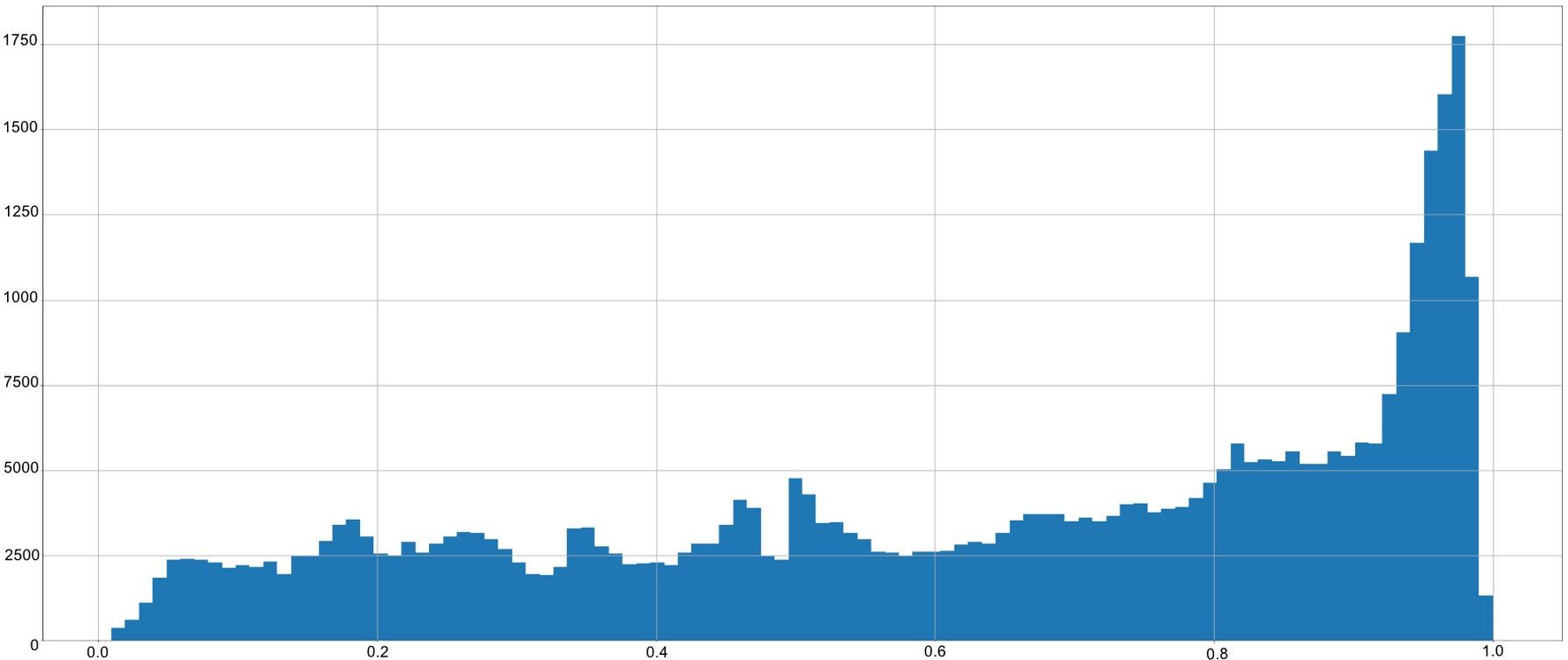}\\
    \includegraphics[width=1\textwidth]{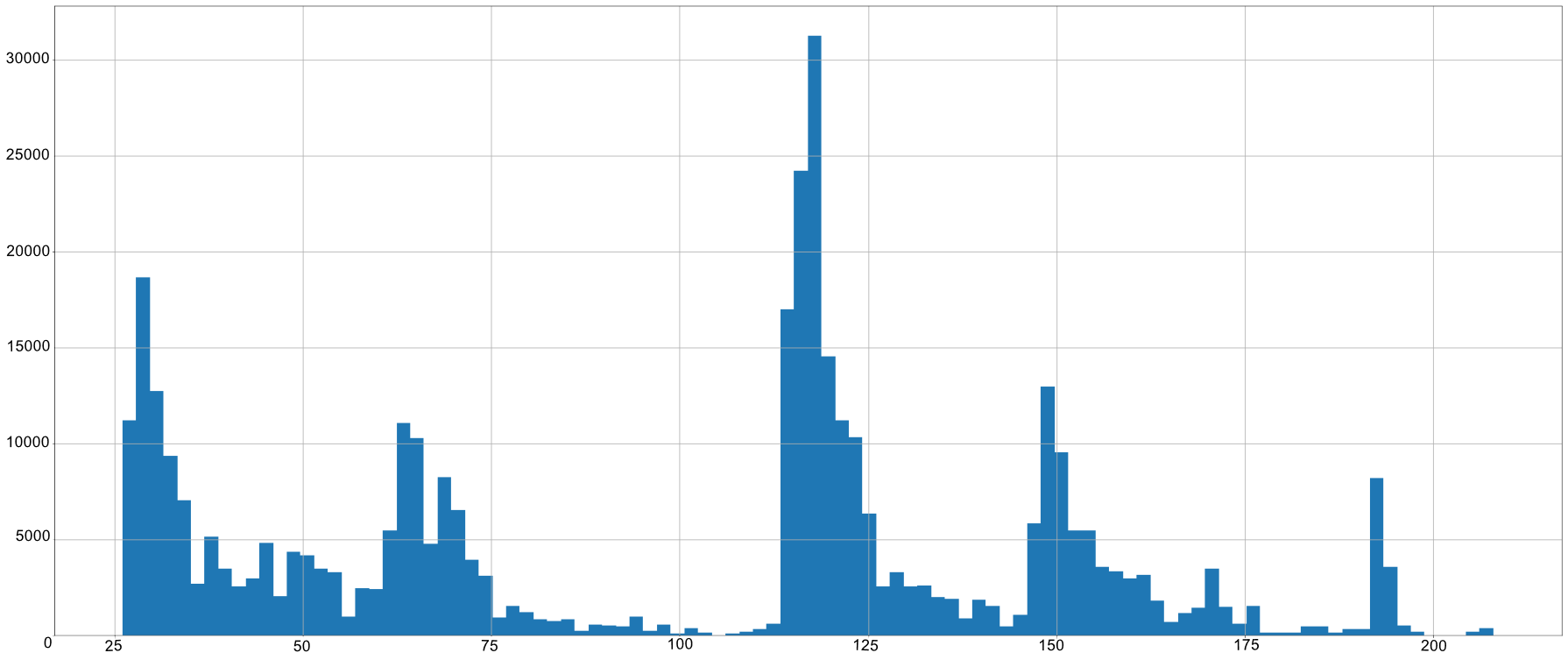}\\
    \includegraphics[width=1\textwidth]{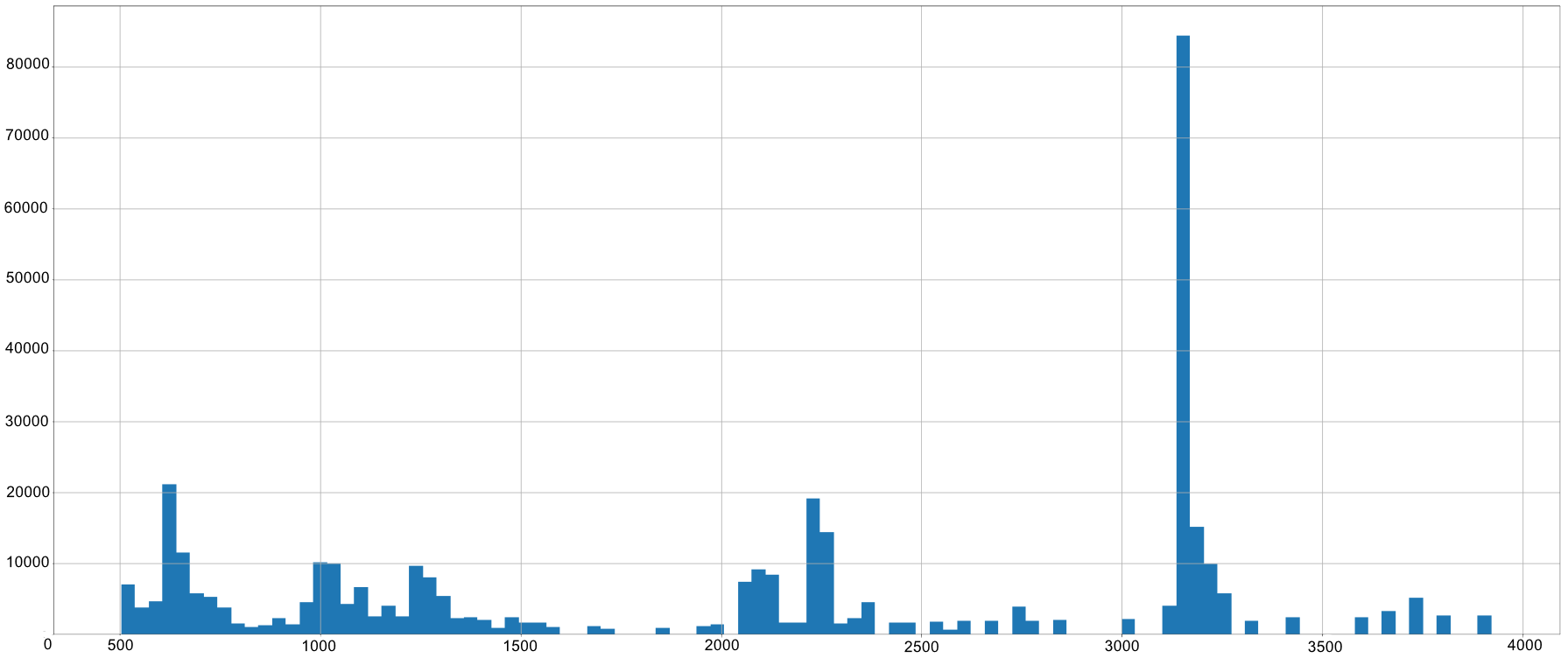}\\

  \caption{{\normalsize The distribution of the number of biclusters by their density (top), extent (middle) and intent sizes (bottom).}}\label{plot1}
\end{figure}

As we can see from the graph in Figure \ref{plot1}, there is a reasonable amount of biclusters with a density value greater than 0.9. The distributions of biclusters by extent and intent show that the majority of biclusters have about 90 samples and 2,600 SNPs, respectively.

\begin{figure}[ht!]
    \centering
    \includegraphics[scale=0.35]{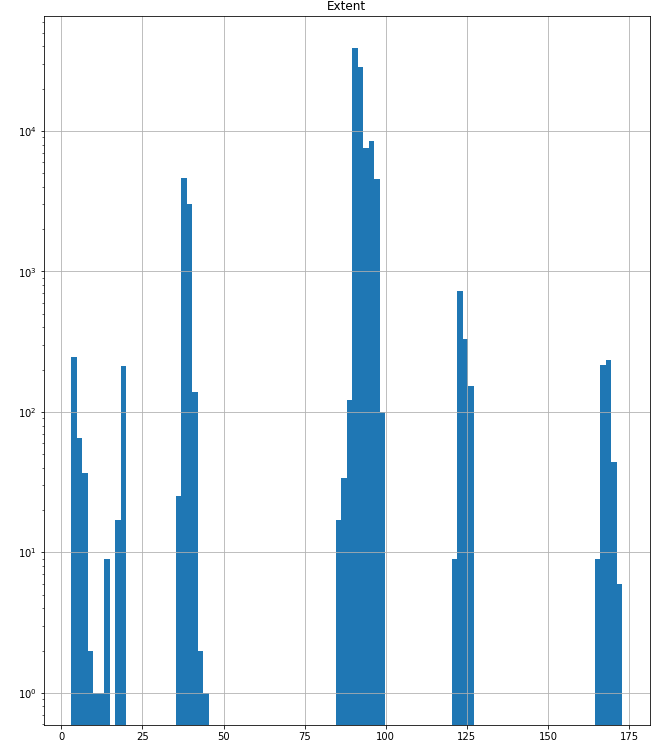} \\
    
    \caption{{\normalsize The distribution of dense biclusters ($\rho_{min}=0.9$) by their extent size.}}\label{plot2}
\end{figure}

For the selection of large dense biclusters, we set the density constraint to be $\rho_{min} = 0.9$. Additional constraints were set as follows: $3 \leq |m'| \leq 1,500$ for the extent size and $3 \leq |g'| \leq 80,000$ for the intent size. In total, we selected 98,529 OA-biclusters with missing values. For this selection, the graph in Fig.~\ref{plot2} shows the selected peaks of large dense biclusters for different extent sizes.\\

\noindent \textbf{Example 1.} Biclusters in the form $(patients, SNPs)$.

\noindent For generating pair $(g,m)=(1102, rs6704827_A)$ we have that 
$$(m',g') \in \sigma_{(3 \leq |A| \leq 1,500) \wedge (3 \leq  |B| \leq 80,000) \wedge (0.9 \leq \rho(A,B) \leq 1)}(\mathcal B), \mbox{ where}$$

\noindent $(m'g')=(\{1101, 1102, \ldots, 1114 \}$, $\{rs10915587_G, rs284267_A, \ldots,$ $rs12171249_A\})$, $\rho(m',g')\approx0.91$, $|m'|=14$ individuals, $|g'|=758$ SNPs, 9,657 pairs out of 10,612 correspond to missing SNP values. 

We studied further large dense biclusters and chose the densest ones with possibly larger sizes of their extents and intents from each of the peaks identified in their distributions, respectively (Fig.~\ref{plot1}). 

Here are some examples of these subsets with their associated graphs. \\

\noindent \textbf{Example 2.}  We can further narrow down the number of patterns in the previous selection  by looking at the distribution of biclusters by their extent size and choosing proper boundaries. Thus, in Fig.~\ref{plot2}, there is the third largest peak of the number of biclusters near the extent size 125.

For the constraints below

$$\rho_{min}=94.08\% \wedge \rho_{max}=100\% \wedge |g'| = 122  \wedge (3 \leq |m'| \leq 80,000)$$

\noindent the large dense bicluster with its intent size of 455 is identified and selected. Such bicluster has a large number of missing genotypes, which are subject to be eliminated later on. \\

\noindent \textbf{Example 3.} 
The selection around the rightmost peak (\textit{see Fig.~\ref{plot2}}) and further refining of the minimal value density  

$$\rho_{min}=95.4\% \wedge \rho_{max}=100\% \wedge 160 \leq |g'| \leq 175  \wedge (3 \leq |m'| \leq 80,000)$$
    
\noindent resulted in the large bicluster with the extent size of 108 and the intent size of 166.

\subsection{Elimination of Large Biclusters with Missing Genotypes}

After applying the proposed biclustering algorithm to the collected dataset, all large biclusters with missing genotypes were identified and eliminated. That resulted in a new data matrix ready for further analysis\footnote{\url{https://github.com/dimachine/OABicGWAS/}}.  We consolidate the evolution of the two datasets before and after removing missing values in Table~\ref{tbl}.

\begin{table}[ht!]
\caption{Basic statistics of the datasets \textbf{before} and \textbf{after} elimination of missing values.}\label{tbl}
\begin{center}
\begin{tabular}{c|cccc|}
\cline{2-5}
\multicolumn{1}{c|}{}  & no.  & no.  & no. & NaNs\\
\multicolumn{1}{c|}{}  & samples & SNPs &  NaNs &  fraction\\
\hline
\multicolumn{1}{|l|}{Before elimination} & 1,223 & 85,142 & 553,430 & 0.49\% \\
\multicolumn{1}{|l|}{After elimination} & 1,472 & 82,690 & 388,052 & 0.31\%\\
\hline
\end{tabular}
\end{center}
\end{table}

As seen from Table~\ref{tbl}, the biclustering algorithm application resulted in improvement in terms of entries corresponding to SNPs with missing genotypes, a fraction of such entries is reduced by ~29.88\%. The total number of biclusters generated before and after eliminating SNP with missing genotypes is 383,733 (with duplicates) and 259,440, respectively. The total amount of time for generating these biclusters before and after deleting missing data is $3433.2$ and $2293.7$ seconds (by Algorithm~\ref{Alg:MVBiA}), respectively. As for online Algorithm~\ref{alg:online}, it has processed the original context (before elimination) in 1.5 seconds, while the post-processing Algorithm~\ref{alg:online-post} for density computation has taken 907 seconds in sequential and 651 seconds in parallel (six cores) modes, respectively. 

\begin{figure}[ht!]
    \centering
    \includegraphics[width=\textwidth]{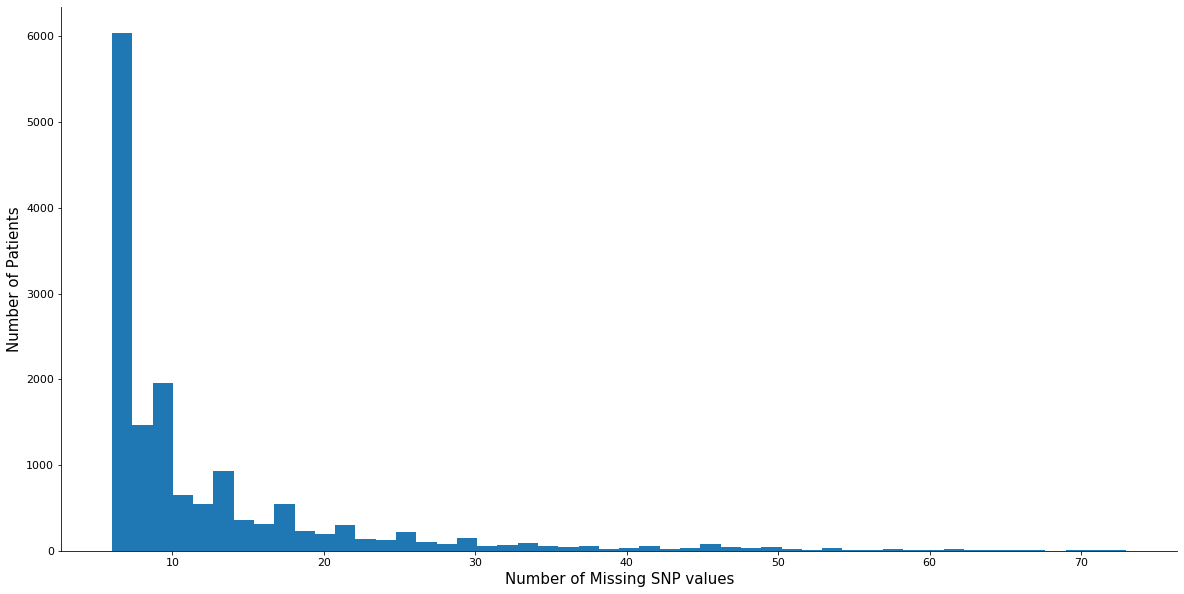} \\
    \caption{{\normalsize Distribution of the number of SNPs with missing genotypes by columns \textbf{after} elimination.}}\label{fig6}
\end{figure}

Fig.~\ref{fig6} shows the distribution of missing values in columns in the new data set (after elimination of missing data), which now has less ragged character.

\subsection{Large Dense Biclusters Elimination and Classification Quality}
We have conducted a number of machine learning experiments on our datasets to check the impact of eliminating missing data. Our proposed algorithm handled on the quality measures of supervised learning algorithms. 

We choose to use gradient boosting on decision trees (GBDT). For this purpose, we selected two libraries where it is already implemented, \textsc{CatBoost} and \textsc{LightGBM}. Both implementations can handle missing values. 

A genome can essentially be interpreted as a sequence of SNPs, so we made a decision to also use \textsc{Long-Short Term Memory Network}~\cite{lstm} as a strong approach to handling sequential data.

\paragraph{\textbf{First dataset experiments.}} Firstly, we applied GBDT algorithm from \textsc{CatBoost} library to our initial dataset (before elimination of SNPs with missing genotypes). The following parameters were taken for the classifier:
\begin{itemize}
    \item Maximum number of trees: 3;
    \item Tree depth limit: 3;
    \item Loss function: binary cross-entropy (log-loss/binary cross-entropy).
\end{itemize}

We also applied LSTM approach the following way: the initial sequence was resized to 100 elements by a fully-connected layer, then the layer output was passed to the LSTM module element-wise. The hidden state of LSTM after the last element was passed to a fully-connected classification layer.

The scores on this dataset were evaluated with 3-fold cross-validation with stratified splits. Basic classification metrics' scores are present in Table \ref{sco}.

\begin{table}[]
    \centering
    \caption{{\normalsize Classification scores on the test set \textbf{before} elimination of missing SNP values}}\label{sco}
    \begin{tabular}{|l|l|l|l|l|}
    \hline
     & \textbf{accuracy} & \textbf{F1-score} & \textbf{precision} & \textbf{recall} \\ \hline
    CatBoostClassifier & 0.966 & 0.9758 & 0.9558 & 0.9967 \\ \hline
    FC+LSTM & 0.890 & 0.926 & 0.880 & 0.982 \\ \hline
    \end{tabular}
    
\end{table}

These unexpectedly high scores were unrealistic since the GBDT model complexity had one of the lowest possible configurations, and the LSTM model, which is handling the data in a different way, also achieved high accuracy. For a lot of samples, the model learned to ``understand'' on which chip it was analyzed by looking at the patterns of missing genotypes, so the data leak was present. 

\paragraph{\textbf{Second dataset experiments.}} This dataset was obtained after the identification of large dense biclusters by application of our proposed algorithm with subsequent elimination. Table~\ref{expre} recaps the experiments conducted on the dataset. For the first and second experiments, we used \textit{CatBoost classifier}  with train/test split in the proportion of 8:2 and 3-fold cross-validation, respectively, while maintaining the balance of classes for model validation. In the third experiment, we used \textit{LGBMClassifier} classifier with 3-fold cross-validation while maintaining the balance of classes for model validation. In the fourth experiment, the described earlier \textit{LSTM} classifier was used with the aforementioned cross-validation.

\begin{table}[ht!]
\caption{{\normalsize Scores results of different machine learning classifiers applied to the dataset \textbf{after} elimination of SNP with missing genotypes.}}\label{expre}
\begin{center}
\begin{tabular}{c|cccccc|}
\cline{2-7}
                                                          & \textbf{no. trees} & \textbf{depth} & \textbf{accuracy} & \textbf{F1-score} & \textbf{precision} & \textbf{recall} \\ \hline
\multicolumn{1}{|c|}{\multirow{3}{*}{CatBoostClassifier}} & 2                 & 2              & 0.715             & 0.834             & 0.715              & \textbf{1.000}  \\
\multicolumn{1}{|c|}{}                                    & 5                 & 2              & \textbf{0.773}    & \textbf{0.862}    & \textbf{0.761}     & 0.995           \\
\multicolumn{1}{|c|}{}                                    & 5                 & 3              & \textbf{0.773}    & \textbf{0.862}    & \textbf{0.761}     & 0.995           \\ \hline
\multicolumn{1}{|c|}{\multirow{2}{*}{CatBoostClassifier}} & 4                 & 3              & \textbf{0.768}    & \textbf{0.859}    & \textbf{0.990}     & \textbf{0.759}  \\
\multicolumn{1}{|c|}{}                                    & 5                 & 3              & \textbf{0.768}    & \textbf{0.859}    & \textbf{0.990}     & \textbf{0.759}  \\ \hline
\multicolumn{1}{|c|}{\multirow{5}{*}{LGBMClassifier}}     & 5                 & 3              & 0.753             & 0.852             & \textbf{0.997}     & 0.744           \\
\multicolumn{1}{|c|}{}                                    & 5                 & 5              & 0.753             & 0.852             & 0.996     & 0.744           \\
\multicolumn{1}{|c|}{}                                    & 4                 & 4              & 0.751             & 0.851             & \textbf{0.997}     & 0.742           \\
\multicolumn{1}{|c|}{}                                    & 4                 & 3              & 0.749             & 0.850             & \textbf{0.997}     & 0.741           \\
\multicolumn{1}{|c|}{}                                    & 5                 & 4              & \textbf{0.756}    & \textbf{0.854}    & 0.996              & \textbf{0.747}  \\ \hline

\multicolumn{1}{|c|}{FC+LSTM}     & -                 & -              & 0.731             & 0.839             & 0.735     & 0.981           \\ \hline

\end{tabular}
\end{center}
\end{table}

From Table \ref{expre}, one can see that scores are more realistic in comparison to those of Table~\ref{sco}, thus showing us that data leak and subsequent overfitting effects are gone. We realize that our proposed biclustering algorithm successfully identified large submatrices with missing data, which we eliminated and successfully removed the impact of data leak and overfitting.

\subsection{Detecting concepts of missing SNP values under size constraints}

In-Close4 is an open-source software tool~\cite{Andrews:2017}, which provides a highly optimised algorithm from \textsc{CbO} family~\cite{kuznetsov:1996,Janostik:2020} to construct the set of concepts satisfying given constraints on sizes of extents and intents. In-Close4 takes as input a context  and outputs a reduced concept lattice: all concepts satisfying the constraints given by parameter values ($|A| \geq m$ and $|B| \geq n$, where $A$ and $B$ are extent and intent of an output formal concept, and $m, n\in \mathbb{N}$). 

To deal with our large real-world dataset, we changed the maximum default values used in the executable of In-Close4 parameters as follows:

\storestyleof{itemize}
\begin{listliketab}
    \begin{tabular}{Llll}
         &  \#define MAX\_CONS 30000000           & //max number of concepts   \\
         &  \#define MAX\_COLS 90000 & //max number of attributes  \\
         & \#define MAX\_ROWS 2000  &  //max number of objects \\
    \end{tabular}
\end{listliketab}
\\

From Tables~\ref{tbl:IncBefore}~and~\ref{tbl:IncAfter}, one can see that the number of concepts generated by In-Close4 becomes several times larger than that of OA-biclusters, in our case study. When we set the extent size constraint to 5 with the input context before and the extent and the intent size constraint to 20 and 0, respectively, after the elimination of missing data, the software crashed. Meanwhile, our proposed biclustering algorithms could manage to output all OA-biclusters in both cases. 

As the author of InClose suggested in private communication, the tool was optimised for ``tall'' contexts with a large number of objects rather than attributes, while in bioinformatics the contexts are often ``wide'' like in our case when the number of SNPs is almost 57 times larger than that of individuals. So, the results on the transposed context along with properly set compilation parameters allowed to process the whole context for $m=0$ and $n=0$\footnote{The last line in Table~\ref{tbl:IncBefore} and the last five lines in Table~\ref{tbl:IncAfter} corresponds to the experiments conducted for the final version of the paper on the transposed contexts.}. 

\begin{table}[ht!]
\caption{The number of concepts and elapsed time generated by In-Close4 algorithm \textbf{before} eliminating SNPs with missing genotypes.}\label{tbl:IncBefore}
\begin{center}
\begin{tabular}{cccc}
\hline
Min intent size & Min extent size & Total Time, s & No. of Concepts \\ \hline
0 & 45 & 21.2 & 18,617 \\
0 & 40 & 23.6 & 34,400 \\
0 & 30 & 35.8 & 68,477 \\
0 & 20 & 46.1 & 165,864 \\
0 & 10 & 64.3 & 214,007 \\
0 & 5 & 188.3 & 1,220,576 \\ 
\hline
0 & 0 & 143.43 & 1,979,439 \\
\hline
\end{tabular}
\end{center}
\end{table}

\begin{table}[ht!]
\caption{{\normalsize The number of concepts and elapsed time generated by In-Close4 algorithm \textbf{after} eliminating SNPs with missing genotypes.}}\label{tbl:IncAfter}
\begin{center}
\begin{tabular}{cccc}
\hline
Min intent size & Min extent size & Total Time, s & Number of Concepts \\ \hline
0 & 40 & 10.4 & 2,743 \\
0 & 30 & 10.6 & 4,196 \\
0 & 20 & 12.6 & 19,620 \\ 
\hline
30 & 0 & 5.8 & 352,257\\
25 & 0 & 6.2 & 466,695\\
20 & 0 & 7.4 & 695,962\\
15 & 0 & 10.7 & 1,308,222\\
10  & 0 & 18.3 & 3,226,277\\
\hline
\end{tabular}
\end{center}
\end{table}

Even if we do not know the number of output concepts for the context after elimination of missing SNP values, their number is more than 10 times larger than that of OA-biclusters, which might be considered as argument in favour of their usage for the studied problem with rather low or no size constraints.

\section{Conclusion}\label{con}

A new approach to process the missing values in datasets of SNP genotypes obtained with DNA-microarrays is proposed. It is based on OA-biclustering. We applied the approach to the real-world datasets representing the genotypes of patients with ischemic stroke and healthy people. It allowed us to estimate and eliminate the SNPs carefully with missing genotypes. Results of the OA-biclustering algorithm showed the possibility of detecting relatively large dense biclusters, which significantly helped in removing the effects of data leaks and overfitting while applying ML algorithms. 

We compared our algorithm with In-Close4. The number of OA-biclusters generated by our algorithm is significantly lower than the number of concepts (or biclusters) generated by In-Close4. Besides, our algorithm has the advantage of using OA-bicluster without the need to experiment with finding the best minimum support, as in the case of using In-Close4 for generating formal concepts. 

Since survey~\cite{Naulaerts:2015} mentioned frequent itemset mining (FIM) as a tool to identify strong associations between allelic combinations associated with diseases, the proposed algorithm needs further comparison with other approaches from FIM like DeBi~\cite{serin2011debi} and anytime discovery approaches like Alpine~\cite{Hu:17} tested on GEA datasets as well; though their use may get complicated if we need to keep information about object names for decision-makers. It also requires further time complexity improvements to increase the scalability and quality of the extensive bicluster finding process for massive datasets. 
 
Another venue for related studies delve in Boolean biclustering~\cite{Michalak:2019} and factorisation techniques~\cite{Belohlavek:2019}.

Speaking about other possible applications of biclustering, we suggest the development of a new imputation technique. Since biclustering has been recently applied to impute the missing values in gene expression data~\cite{chowdhury2020ncbi} and both GED and SNP genotyping data are obtained with DNA-microarrays and represented as an integer matrix, it can be potentially applied to impute the genotypes that facilitates statistical analyses and empowers ML algorithms. 

\subsubsection*{Acknowledgements.}
This study was implemented in the  Basic Research Program's framework at the National Research University Higher School of Economics and the Laboratory of Models and Methods of Computational Pragmatics in 2020. The authors thank prof. Alexei Fedorov (University of Toledo College of Medicine, Ohio, USA) and prof. Svetlana Limborska (Institute of Molecular Genetics of National Research Centre ``Kurchatov Institute'', Moscow, Russia) for insightful discussions of the results obtained, and anonymous reviewers. 

\subsubsection*{Funding.}
The study was funded by RFBR (Russian Foundation for Basic Research) according to the research project No 19-29-01151. The foundation had no role in study design, data collection and analysis, writing the manuscript, and decision to publish.

\bibliographystyle{splncs}

\bibliography{bib}

\end{document}